\pdfoutput=1
\documentclass[a4paper,12pt]{article}
\title{Non-coherent transport in serial/parallel quantum dots in self-consistent field regime}
\author{S. Illera\\
A. Cirera \\
\small MIND/IN$^2$UB Departament d'Electr\`onica, Universitat de Barcelona,\\\small C/Mart\'i i Franqu\`es 1, E-08028 Barcelona, Spain }
\usepackage{graphicx}
\date{}

\begin{document}
\maketitle

\hrulefill
\begin{abstract}
In this paper we introduce a simple phenomenological model of the conduction between a couple of serial or parallel quantum dots. This model is extended to an arbitrary of number and to a square array of quantum dots. The local potential is computed taking into account the net charge at the quantum dot. Master equations are presented for total current and charge number N considering both the local potential and the local current. Interesting results are reported, namely the Negative Differential Resistance for serial quantum dots as well as resonant conductance. In parallel configuration two independent conduction channels have been observed.\\

\end{abstract}
\hrulefill\\
\small Contact author: sillera@el.ub.es

\section{Introduction}
Single-electron devices are currently conceived to take advantage of the tunnel current between quantum states belonging to nano-scale particles \cite{1,2}. Up to now, the study of such tunnel current has been carried out using simple phenomenological (i.e. rate equation) approaches \cite{3,4} or more complex quantum theories including NEGFF \cite{5,6,7}, transfer Hamiltonian approach \cite{8,9}. Concerning the first case, simple rate equations give us useful and intuitive approaches. The effects of local potential due to selfcharge can be also included by means of a couple of master equations \cite{4}.On the other hand, quantum study of non-equilibrium schemes results in a more complex approach.\\
Until now research has been mostly concentrated on single dots, but rapid progress in microfabrication technology has made possible the extension to couple-dot system with aligned levels \cite{28,24,29}.
A very interesting result was obtained by Shanguan et al \cite{10}. These authors describe the transport in one-dimensional quantum dot array by using many body approaches. Up to now the only computation of transport in an array of quantum dots was done by Carreras et al \cite{11} but no local potential due to selfcharge was included. \\
In this work we present a semiclassical approach based on non-coherent rate equations \cite{4,rate} that is extended from two to an arbitrary number of quantum dots either serial, parallel and square array. The whole array of quantum dots has been treated as a separated quantum system. Then the leads (reservoirs) were incorporated through the rate equations for single dots. In this approach, the coherent effects does not appear and the probabilities of finding an electron in two different dots are coupled by some rate $\gamma$. This treatment is called classical (non-coherent) approach.\\
 The effects of the local potential are computed within the selfconsistent field (SCF) regime. The SCF method is widely used because the exact method based on a multielectron picture is usually impossible to implement. As example we compute the total current and charge transport distribution $N_i$ for two parallel and serial quantum dots (figure 1). The model has been implemented in a home made program in MATLAB code that gives us the current voltage characteristics, the differential conductance and the population inside each dot for a range of applied voltage. Some known results as Negative Differential Resistance (NDR) and resonant conductance are well explained. Moreover we found that the parallel quantum dots act like two independent conduction channels.\\
\begin{figure}[h!]
\centering
\includegraphics[width=0.35\textwidth]{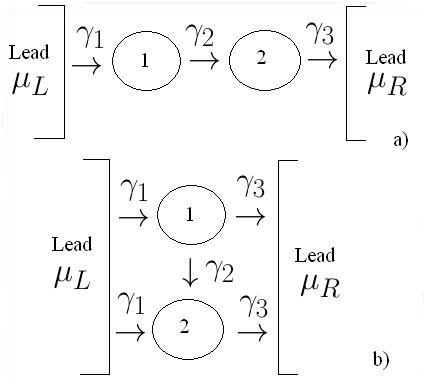}
 \caption{System composed of quantum dot 1 and 2, and two leads. $\gamma_i$ are the tunneling rates between each part of the system. $\mu_R$ and $\mu_L$ are the electrochemical potentials in right and left lead, respectively. Representation of the two different systems: serial configuration (a) and parallel configuration (b).}
\end{figure}

\section{Theroretical model}
When we have two leads coupled to a system with no voltage applied between them, the system is in equilibrium with a common electrochemical potential. In this equilibrium state, the average number of electrons in any energy level is given by the Fermi function:
\begin{equation}
f(E) =\frac{1}{1 + exp((E-\mu)/k_{B}T)}
\end{equation}
An applied bias voltage $V_G = \frac{\mu_L-\mu_R}{e}$ drives the  system out of equilibrium. Each contact tries to bring the system into equilibrium with itself. The system is thus forced into a balancing act between two reservoirs with different population. Then, the average number of electrons $N$ inside the system will be something between $f_R$ and $f_L$.  Rate-equation calculation consider the dynamic behavior of the number of electron $n_i$ at each quantum dot balanced with the incoming and outcoming flux either from a partially filled ($f_{L,R}$) lead or other quantum dot ($n_j$)  \cite{12,13}.\\

\subsection{Serial quantum dot}
Rate-equation for serial Qd can be written as follow: 
\begin{equation}
\frac{dn_1}{dt}=\gamma_1f_L+\gamma_2n_2-(\gamma_1+\gamma_2)n_1
\end{equation}
\begin{equation}
\frac{dn_2}{dt}=\gamma_3f_R+\gamma_2n_1-(\gamma_2+\gamma_3)n_2
\end{equation}
where coefficients $\gamma_i$ are the transition probabilities as described in figure 1a. $f_L$ and $f_R$ are the Fermi-Dirac distribution functions in the electrodes. Density of states at electrodes is considered constant in the energy range that transport takes place. $n_i$ is the average number of electrons at $i-th$ Qd.\\
The steady state solution of (2) and (3) give as the expression 
\begin{equation}
n_1=\frac{\gamma_1(\gamma_2+\gamma_3)f_L+\gamma_2\gamma_3f_R}{(\gamma_1+\gamma_2)(\gamma_2+\gamma_3)-\gamma_2^2}
\end{equation}
\begin{equation}
n_2=\frac{\gamma_3(\gamma_2+\gamma_1)f_R+\gamma_2\gamma_1f_L}{(\gamma_1+\gamma_2)(\gamma_2+\gamma_3)-\gamma_2^2}
\end{equation}
Besides the out coming net flux from Qd2 to the right lead is 
\begin{equation}
I=(-q)\frac{\gamma_3}{\hbar}(f_R-n_2)
\end{equation}
That help us to consider a n-independent way to express the current using (4) and (5) 
\begin{equation}
I^s=\frac{\gamma_1\gamma_2\gamma_3}{\gamma_1\gamma_2+\gamma_1\gamma_3+\gamma_2\gamma_3}(f_L-f_R)
\end{equation}
The expression of the current can be physically interpreted quite easily: the three barriers in the Qd stack act as a series conection of resistors corresponding to the inverse of tunnelling rates $R_{total}=\sum_i^N {R_i}$.

\subsection{Parallel quantum dot}
We can propose the same procedure for parallel Qd. In this case rate-equations are 
\begin{equation}
\frac{dn_1}{dt}=\gamma_1f_L+\gamma_3f_R+\gamma_2n_2-(\gamma_1+\gamma_2+\gamma_3)n_1
\end{equation}
\begin{equation}
\frac{dn_2}{dt}=\gamma_1f_L+\gamma_3f_R+\gamma_2n_1-(\gamma_1+\gamma_2+\gamma_3)n_2
\end{equation}
considering the figure 1b. The steady-state solution of (8) and (9) is 
\begin{equation}
n_1=\frac{\gamma_1(\gamma_1+\gamma_3+2\gamma_2)f_L+\gamma_3(\gamma_1+\gamma_3+2\gamma_2)f_R}{(\gamma_1+\gamma_3)^2+2\gamma_2\gamma_3+2\gamma_2\gamma_1}
\end{equation}
\begin{equation}
n_2=\frac{\gamma_1(\gamma_1+\gamma_3+2\gamma_2)f_L+\gamma_3(\gamma_1+\gamma_3+2\gamma_2)f_R}{(\gamma_1+\gamma_3)^2+2\gamma_2\gamma_3+2\gamma_2\gamma_1}
\end{equation}
The net out coming flow is  
\begin{equation}
I^p=(-q)\frac{\gamma_3}{\hbar}(f_R-n_i)
\end{equation}
for each quantum dot. And substituting (10) and (11) in (12) we obtain the n-independent expression of the current 
\begin{equation}
I^p=\frac{\gamma_1\gamma_3(\gamma_1+\gamma_3+2\gamma_2)}{(\gamma_1+\gamma_3)^2+2\gamma_3\gamma_2+2\gamma_2\gamma_1}(f_L-f_R)
\end{equation}

\subsection{Effect of broadening}
Up to now we are considering only one energy level per dot. But we have missed the broadening of the level that inevitably accompanies any process of coupling to it. The standard way to introduce energy level broadening as a consequence of contacts is to assign a Lorentzian shape to density of states \cite{14,15} 
\begin{equation}
\rho(E)_i=\frac{\frac{\gamma}{2\pi}}{(E-\epsilon_i)^2+(\frac{\gamma}{2})^2}
\end{equation}
The broadening $\gamma$ is proportional to the strength of the coupling. $\gamma=\sum_i\gamma_i$ ,where $\gamma_i$ are the transition rates introduced before. This comes out of a full quantum mechanical treatment, but we could rationalize it as a consequence of the "uncertainty principle" that requires the product of the lifetime $(=\hbar/\gamma)$ of a state and its spread in energy $(\gamma)$ to equal $\hbar$.
The current expression is modified into:
\begin{equation}
I=(-q)\frac{\gamma_1}{\hbar}(f_R-n_1) \rightarrow I=(-q)\int \frac{\gamma_1}{\hbar}\rho_R \rho_1(E)(f_R-n_1)dE
\end{equation}
where now $n_i$ can be interpreted as the non-equilibrium distribution functions. This allow us to compute the total number of electrons at each Qd as  
\begin{equation}
N_i=\int_{-\infty}^{+\infty}dE\rho_in_i\,\,\,\,\,i=1,2
\end{equation}
This procedure allow to rewrite n-independent equations as 
\begin{equation}
I^s=\frac{q}{\hbar}\int_{-\infty}^{+\infty}dE \frac{\gamma_1\gamma_2\gamma_3 \rho_R\rho_L\rho_1\rho_2}{\gamma_1\gamma_2\rho_L\rho_1+\gamma_1\gamma_3\rho_R\rho_L+\gamma_2\gamma_3\rho_2\rho_L}(f_L-f_R)
\end{equation}
\begin{equation}
\small{I_i^p\!\!\!=\!\!\!\frac{q}{\hbar}\!\!\int_{-\infty}^{+\infty}\!\!\!\!\!\!\!\!\!\!\!dE\!\!\frac{\gamma_1\gamma_3(\gamma_1\rho_L+\gamma_3\rho_R+\gamma_2(\rho_1+\rho_2))\rho_L\rho_R\rho_i}{(\gamma_1\rho_L\!\!\!+\!\gamma_3\rho_R)^2\!\!+\!\!\gamma_3\gamma_2\rho_R(\rho_1\!\!+\!\!\rho_2)\!\!+\!\!\gamma_1\gamma_2\rho_L(\rho_1\!\!+\!\!\rho_2)}\!\!(\!f_L\!\!-\!\!f_R\!)}
\end{equation}
for serial and parallel currents. Where $\rho_R$ and $\rho_L$ are the density of states  in the leads (constants), while $\rho_1$ and $\rho_2$ are the DOS in each Qd.

\section{Local potential}
Another factor that we have to take into account is how the voltage applied to the external electrodes change the electrostatic potential inside each dot. It is easy to see that this can play an important role to determining the shape of the current-voltage characteristics. The classical right solution of potential at each quantum dot involves Poisson equation:
\begin{equation}
\vec{\nabla} (\epsilon_r  \vec{\nabla} V)=-\frac{\triangle \rho}{\epsilon_0}
\end{equation}
Where solution is 
\begin{equation}
U=U_L+\frac{q^2}{C_{tot}} \triangle N
\end{equation}
Being $U_L$ the Laplace solution of the system. Consi\-dering the capacitive coupling and boundary conditions 
\begin{equation}
U_L=\sum_i \frac{C_i}{C_{tot}}(-qV_i)
\end{equation}
Where $i$ covers all the elements of the system, including electrodes and arbitrary number of Qd, and $C_i$ is the capacitive coupling between $i$ and $i+1$ element. Finally $C_{tot}=\sum_iC_i$. The charge energy system constant $U_0=q^2/C_{tot}$ gives us the potential increase as a consequence of the electron addition. We observe that potential depends on the increasing charge density, but at the same time charge density depends on energy level, a function of local potential. This considerations confers equation for N (16) and for U (23), the position of master equations, as usual in similar cases (5) and (6).\\
For serial Qd we can consider that first Qd is capacitive coupled with left lead and second Qd, while the  
second Qd is only coupled with the first Qd and the right lead. Thus the Laplace  potential results are 
\begin{eqnarray}
U_{L1}^S=\frac{C_s}{C_{tot}}(-q V_G)+\frac{C_c}{C_{tot}} U_2 \\
U_{L2}^S=\frac{C_c}{C_{tot}} U_1
\end{eqnarray}
where we consider $V_G$ the potential of left electrode and 0 for the right one. $C_s$ and $C_c$ are coupling capacities between Qd and electrode and between Qds. $U_i$ is the total potential in the $i-th$ Qd.\\
Similar expressions can be derived for the parallel system: 
\begin{eqnarray}
U_{L1}^P=\frac{C_s}{C_{tot}}(-q V_G)+\frac{C_c}{C_{tot}} U_2\\
U_{L2}^P=\frac{C_s}{C_{tot}}(-q V_G)+\frac{C_c}{C_{tot}} U_1
\end{eqnarray}
We shall solve $U_1$ and $U_2$ equations simultaneously in an iterative process forcing comparison of results \cite{16} and thus ensuring simultaneous convergence.

\section{Results}

For simplicity we consider fully symmetric system $\gamma_i=\gamma_j$ and $\epsilon_i=\epsilon_j$. The effects of local potential on DOS should be computed as $\rho(E)\rightarrow \rho(E-U)$, which will modify computations for population $N$ and currents $I$. 
Transport requires energy levels of Qds to be between electrochemical potentials of leads. Moreover overlapping of the DOS is also required. Both conditions ensure electron transport and net current. 
In this model we consider $\gamma_i<<U_0$ and $K_BT<<U_0$. Both conditions are necessary for Coulomb Blockade \cite{17}. The self-consistent charging model based on the Poisson equation represents a good zero-order approximation (called Hartree approximation) to the problem of electron-electron interactions, but it is generally recognized that it tends to overestimate the effect. Corrections for the so-called exchange and correlation effects are often added, but the description is still within the one-electron picture which assumes that a typical electron feels some average potential, U, due to the other electrons.  Despite our self-consistent potential model do not explain potential differences between spin up and down as consequence of broken spin degeneration \cite{18,19}, we can obtain a first approach to Coulomb Blockade.

Figures 2a and 2b show obtained intensities for parallel model in a range of capacities values. The $I(V)$ curve is strongly dependent on $C_s$ and $C_c$ through the Laplace
\begin{figure}[h!]
\centering
\includegraphics[width=0.50\textwidth]{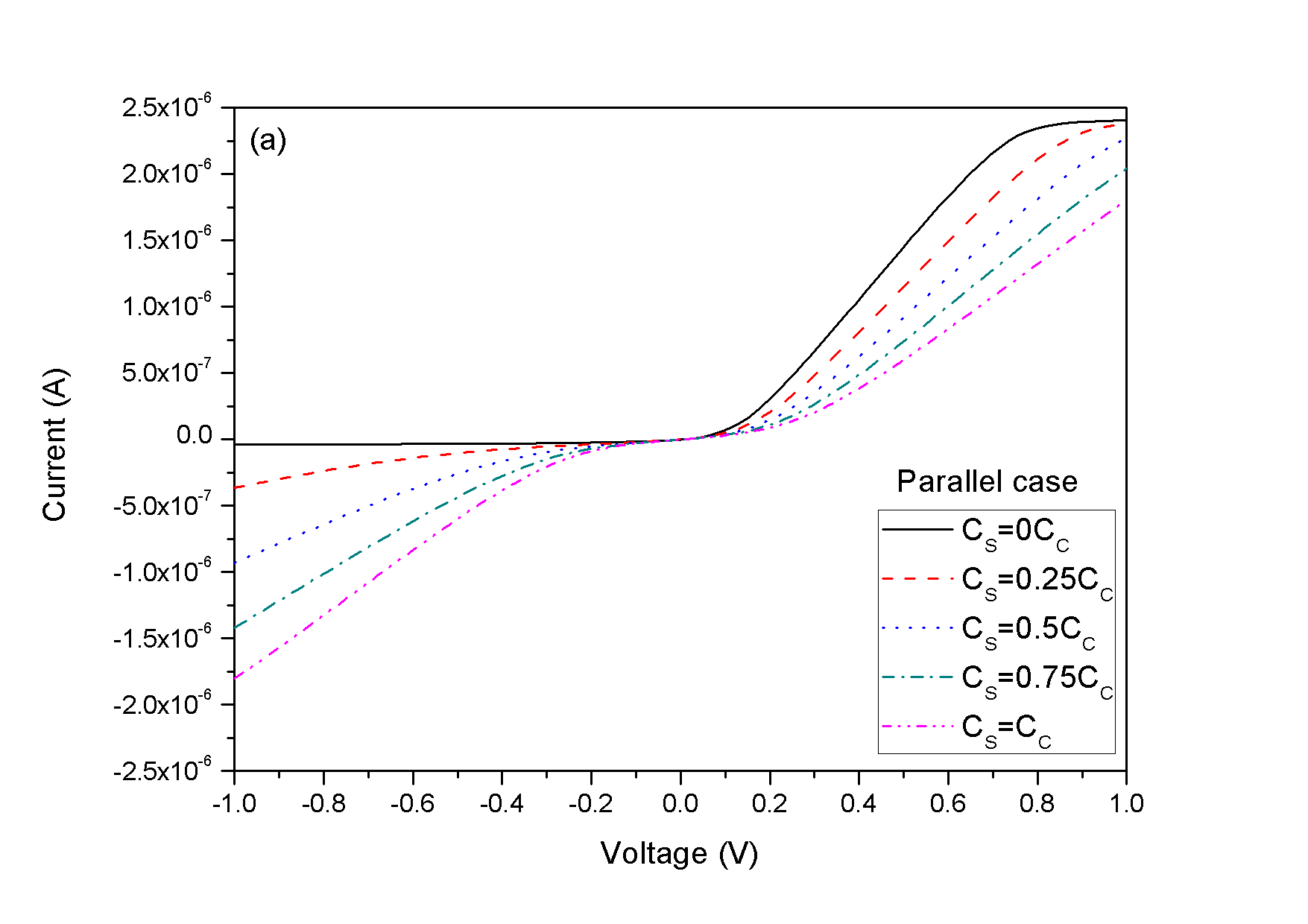}
\includegraphics[width=0.50\textwidth]{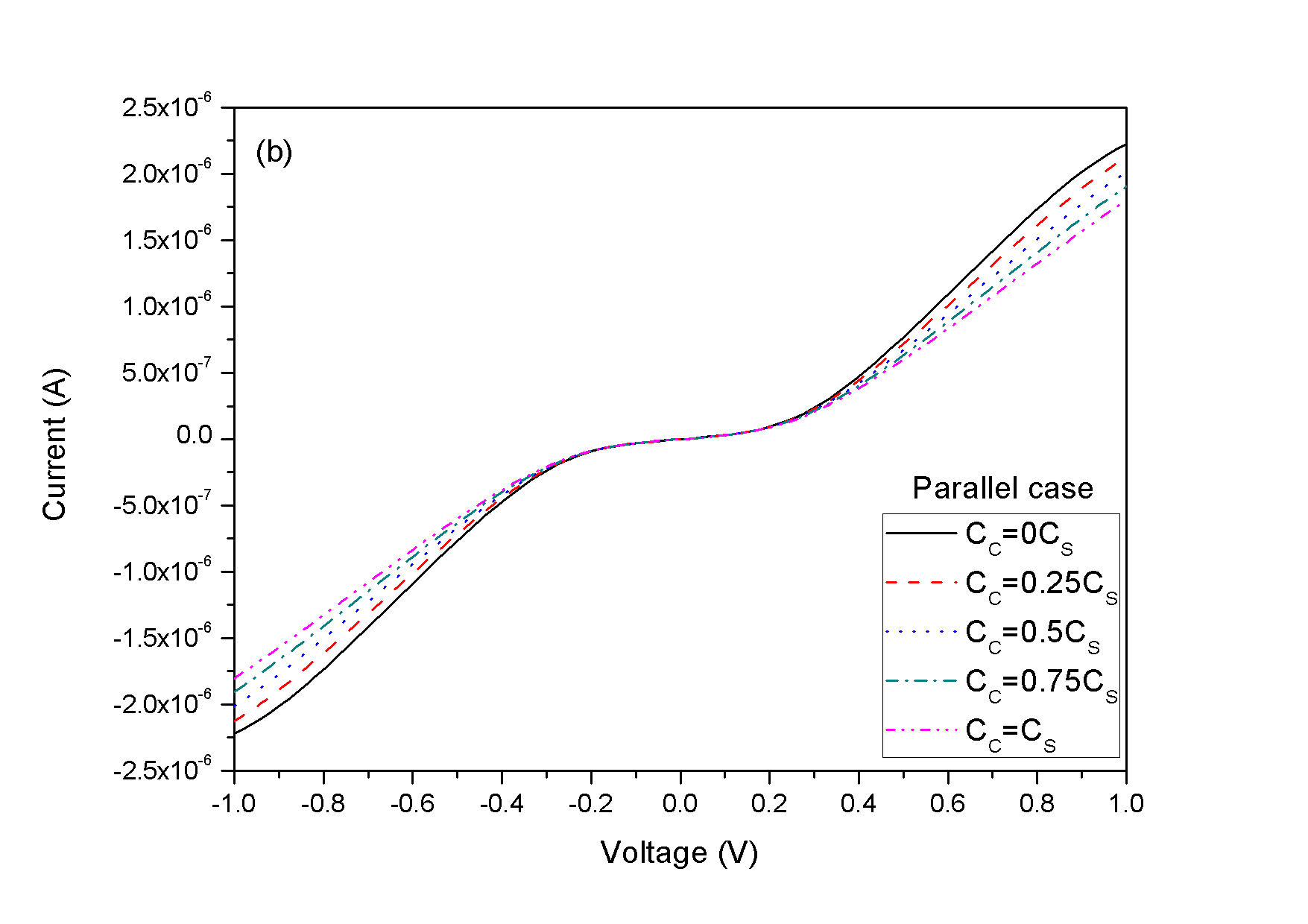}
 \caption{$I(V)$ characteristics calculated in the parallel configuration in a different range of parameters $C_C$ and $C_S$. $C_{s2}$ is a constant . In the upper figure (a) a rectifying effect is obtained due to the weak coupling between the Qds and the leads. In (b) the effect of the electrostatic coupling between the two Qds is shown. $\mu_R=0$,$\gamma_1=\gamma_2=\gamma_3=0.005eV$, $K_BT=0.025eV$, $U_0=0.25eV$,  $\epsilon_1=0.2eV$, $\epsilon_2=0.2eV$.}
\end{figure}
solution of the system. The value of the capacity indicates how the system is coupled. When $C_s$ tends to cero, the two Qd are electrically decoupled with the leads and the potential has not Laplace term, this case implies a lightly coupled system. The other case is when the value of the capacity between Qd ($C_c$) goes to cero, which means no electrical influence between Qd. The Poisson term follows the charge $N_i$ at each Qd and it is therefore always positive, inducing a shift on the potential.
Conductance is depicted in figures 3a and 3b for the two ranges of capacities. In coherence with results shown in \cite{18} conduction gap is found. The peak of conduction becomes finite as consequence of the temperature. Furthermore the charge effects tend to shift and widening the peak.

\begin{figure}[h!]
\centering
\includegraphics[width=0.50\textwidth]{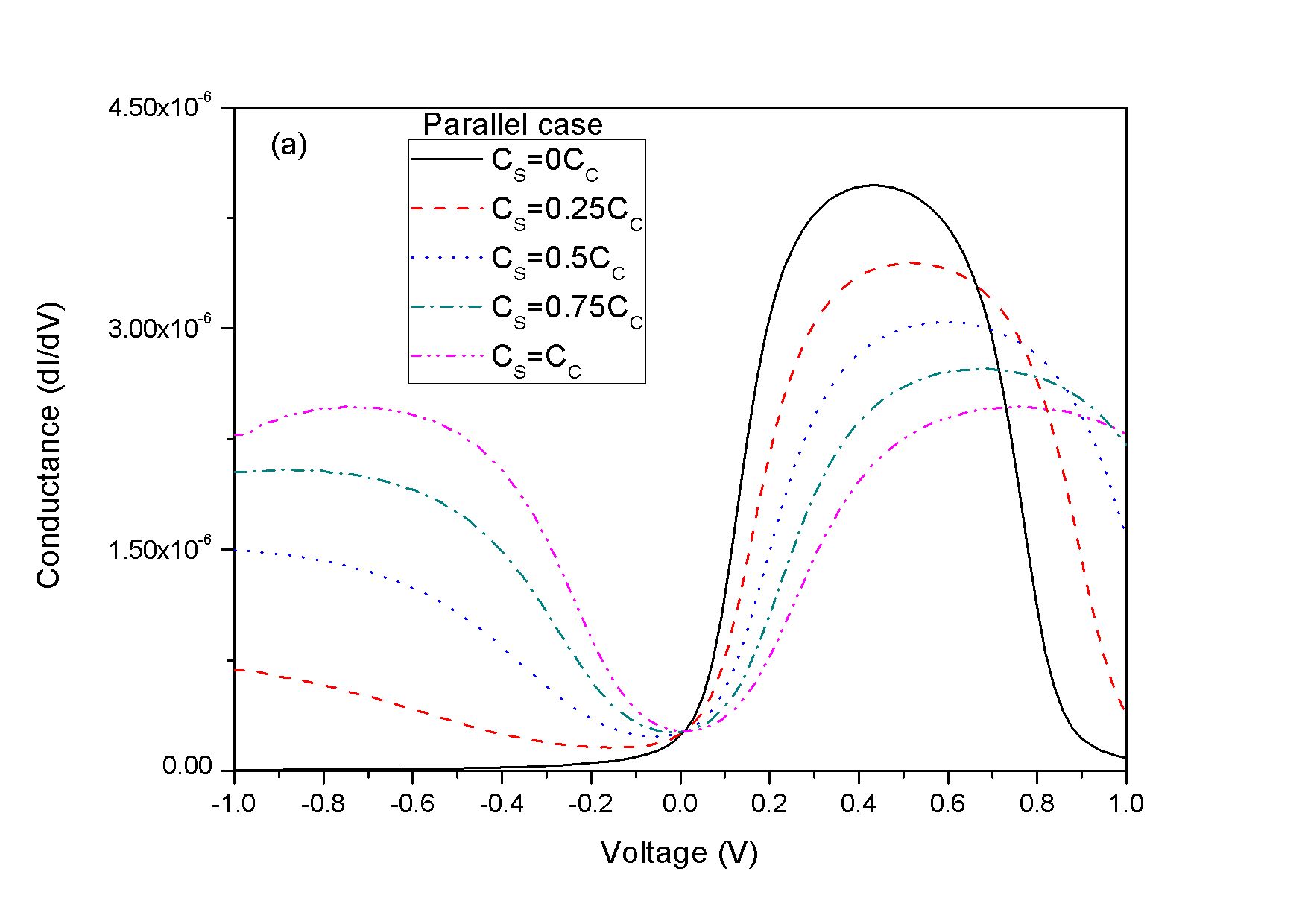}
\includegraphics[width=0.50\textwidth]{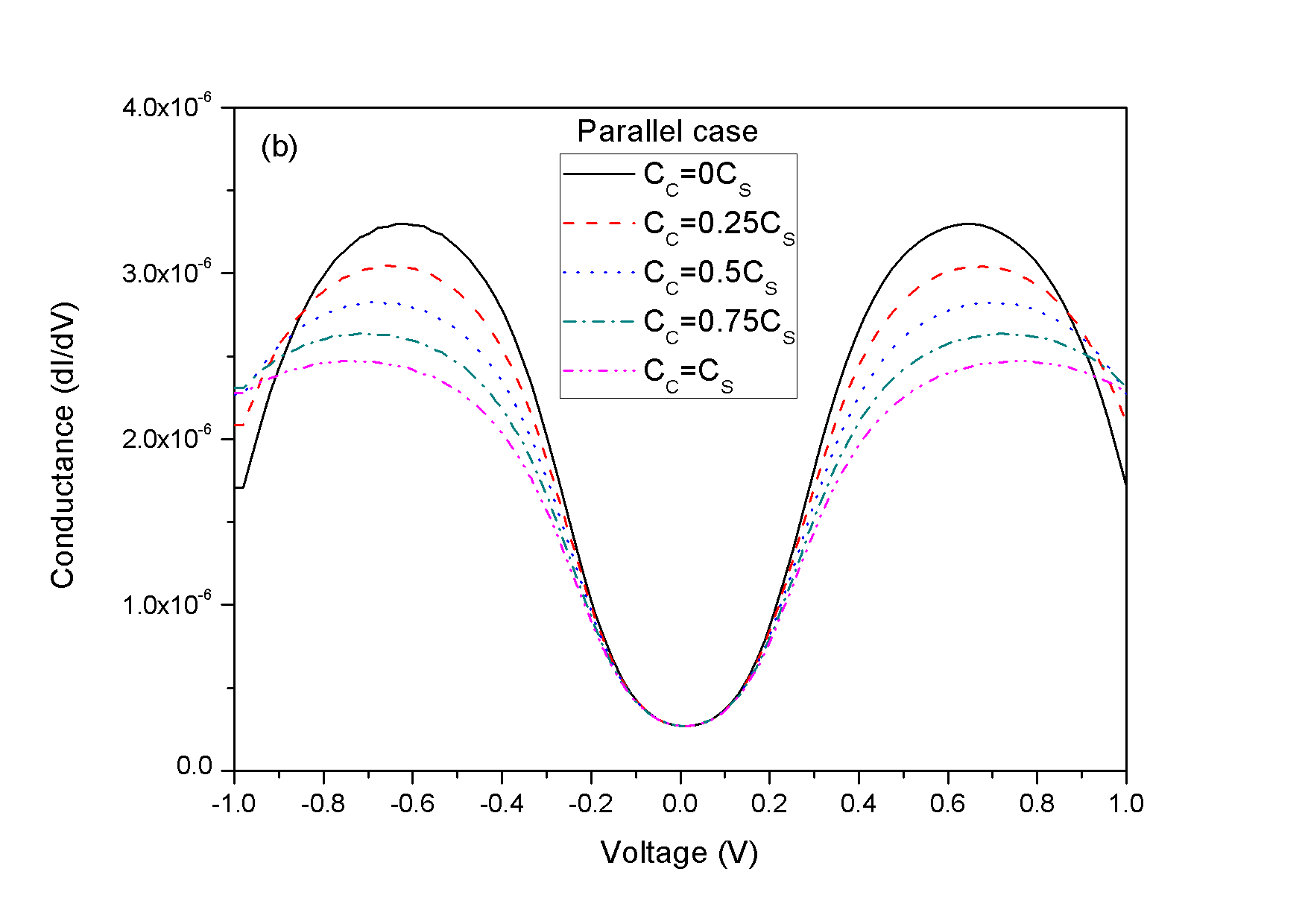}
 \caption{Conductance characteristics obtained in the parallel case, in the same cases as before. (a) Resonant conductance peak appears as a result of the rectifying effect. When $C_S$ increases the peak diminishes and the rectifying effect disappears. (b) Conductance gap is obtained .}
\end{figure}

Concerning the symmetry let us take into account two considerations. First we compute the Laplace domain at each Qd, as follow 
\begin{equation}
\epsilon_1\!\!=\!\epsilon_0\!+\!\frac{C_s}{C_{tot}}(-qV_G\!)\!+\!\frac{C_c}{C_{tot}}U_{2}\,\,\, \mbox{where}\,\, U_{2}\!=\!\frac{C_s}{C_{tot}}(-qV_G)\!+\!\frac{C_c}{C_{tot}}U_{1}
\end{equation}
that define a recurrence, (24) and (25). In order to analyze the results we will extend the series for the energy level of one dot. Thus energy level becomes 
\begin{equation}
\epsilon_1=\epsilon_0+\frac{C_s}{C_{tot}}(-qV_G)+\frac{C_c}{C_{tot}}\frac{C_s}{C_{tot}}(-qV_G)+\frac{C_c^2}{C_{tot}^2}U_1 \approx \epsilon_0+\frac{C_s}{C_s+C_{s2}}(-qV_G)
\end{equation}
Where we have used that the recurrence is a geometric series, $a+ar+ar^2+\ldots=\frac{a}{1-r}$. The value $C_{s2}$ is the coupling capacity with the second lead. As previously stated, the necessary condition for transport is energy level must stay between the electrochemical potentials $\mu_1>\epsilon_i>\mu_2$ of the lead’s. In parallel 
\begin{figure}[h!]
\centering
\includegraphics[width=0.5\textwidth]{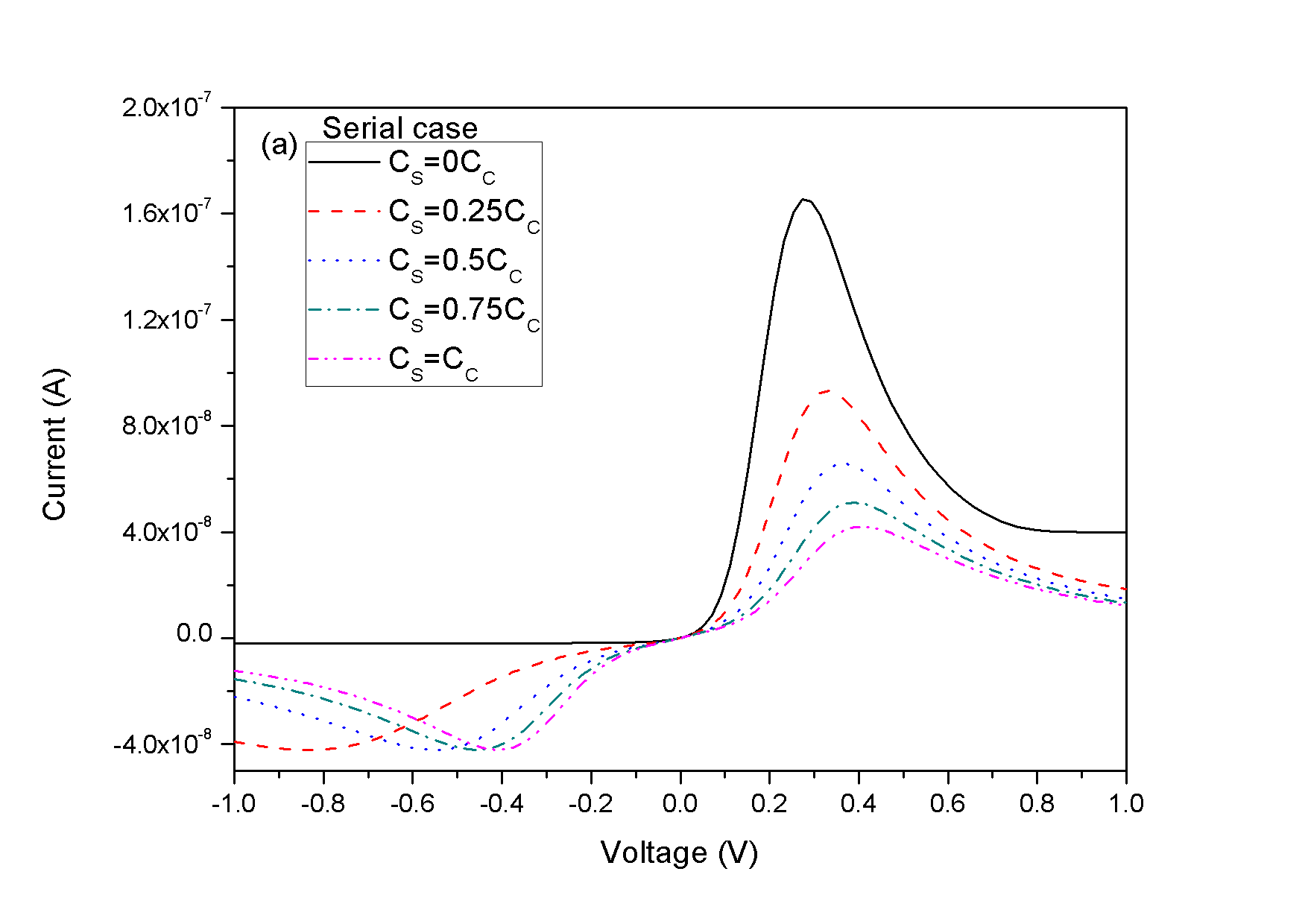}
\hspace{1cm}
\includegraphics[width=0.5\textwidth]{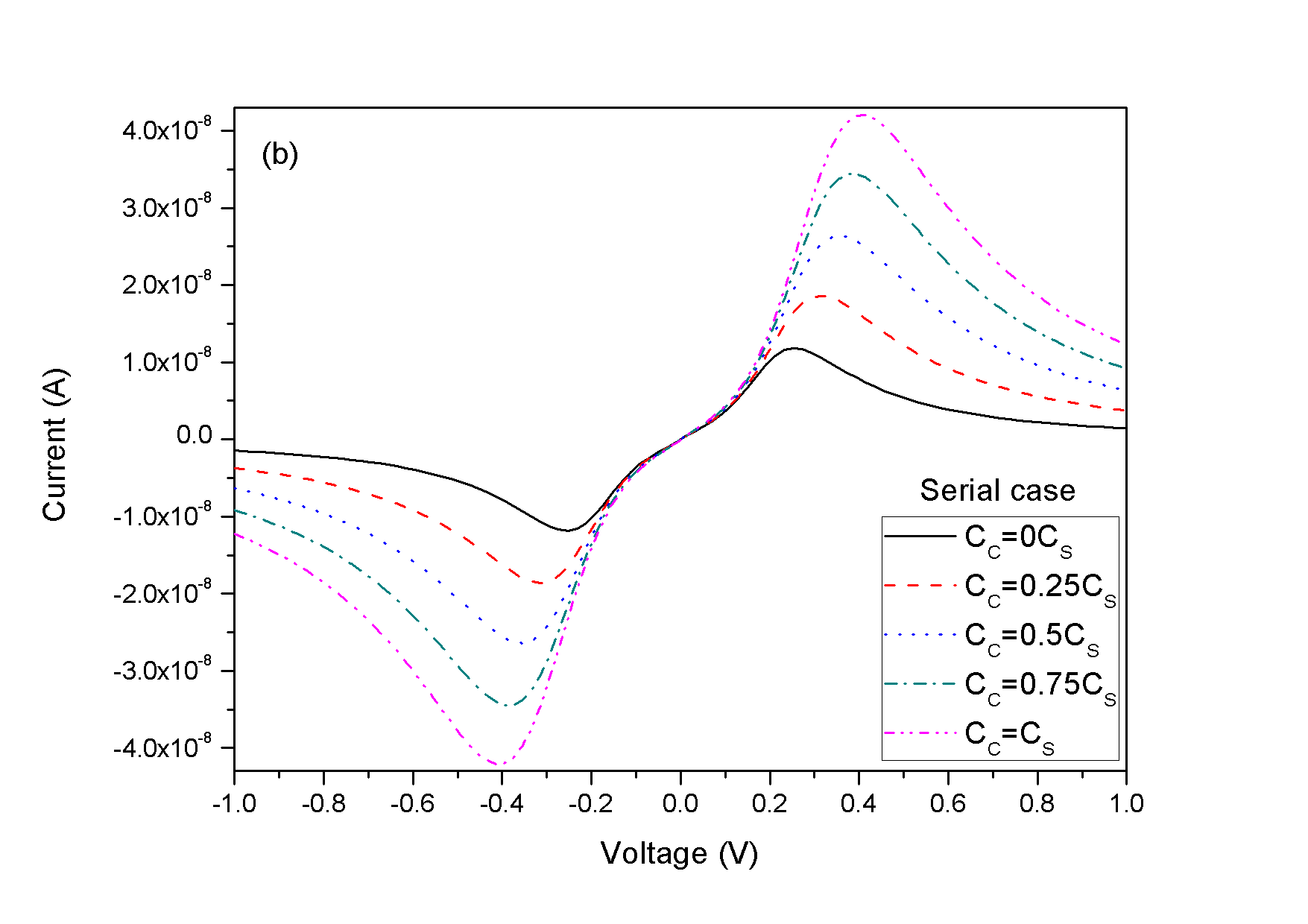}
\vspace{-0.5cm}
 \caption{(a) $I(V)$ curve of  the  double dot in serial configuration, showing rectification effect when it is weakly coupled with the lead and the two levels are aligned. The rectifying effect disappears and resonant peaks appears when the Qd is strongly coupled with the lead. (b) Resonant peaks obtained due to the different electrostatic coupling in each dot. When the voltage increases, the separation between the energy levels increases and the overlapping between the DOS decreases, as a consequence, a NDR appears.  $\mu_R=0$,$\gamma_1=\gamma_2=\gamma_3=0.005eV$, $K_BT=0.025eV$, $U_0=0.25eV$,  $\epsilon_1=0.2eV$, $\epsilon_2=0.2eV$.}
\end{figure}
Qd system, there are multiple pathways possible for electron transport and thus the condition $\epsilon_1=\epsilon_2$ is not required. Then, transport starts when lead electrochemical potential reach the minimum energy level $\epsilon_i$. The potential threshold becomes:
\begin{eqnarray}
V_G^+\!\!=\!\!\frac{\epsilon_0}{1-\frac{C_s}{C_s+C_{s2}}}\\
V_G^-\!\!=\!\!-\frac{\epsilon_0}{\frac{C_s}{C_s+C_{s2}}}
\end{eqnarray}
Two potential thresholds appear, for positive and ne\-gative voltage, as a consequence of previous condition. This creates a region in which no transport occurs because the previous condition is not accomplished, hence a conductance gap appears. The symmetry of the $I(V)$ curves is reached when $V_G^+=\mid \!\!V_G^-\!\!\mid$. This condition occurs when $\epsilon_0=0$ or $C_s=C_{s2}$. In the case of a non-degenerate semiconductor the only possibility of broken symmetry is the different capacitive coupling between the two leads. As we can see from the expression (27) and its equivalent for $\epsilon_2$, if the two Qds have the same $\epsilon_0$ and equal capacity values each Qd acts like a independent conduction channel due to the electrostatic potential is equal in each dot. Furthermore the electrostatic coupling between neighbours is a second order effect.  \\

Figures 4a and 4b show obtained intensities for serial model in a range of capacity values, in the same cases as before. Now, the electron has single transport way and conditions for transport are more restrictive. Overlapping of the density of states between Qds is also required. Conductance is depicted in figure 5a and 5b for the two ranges of capacities. 

\begin{figure}[h!]
\centering
\includegraphics[width=0.5\textwidth]{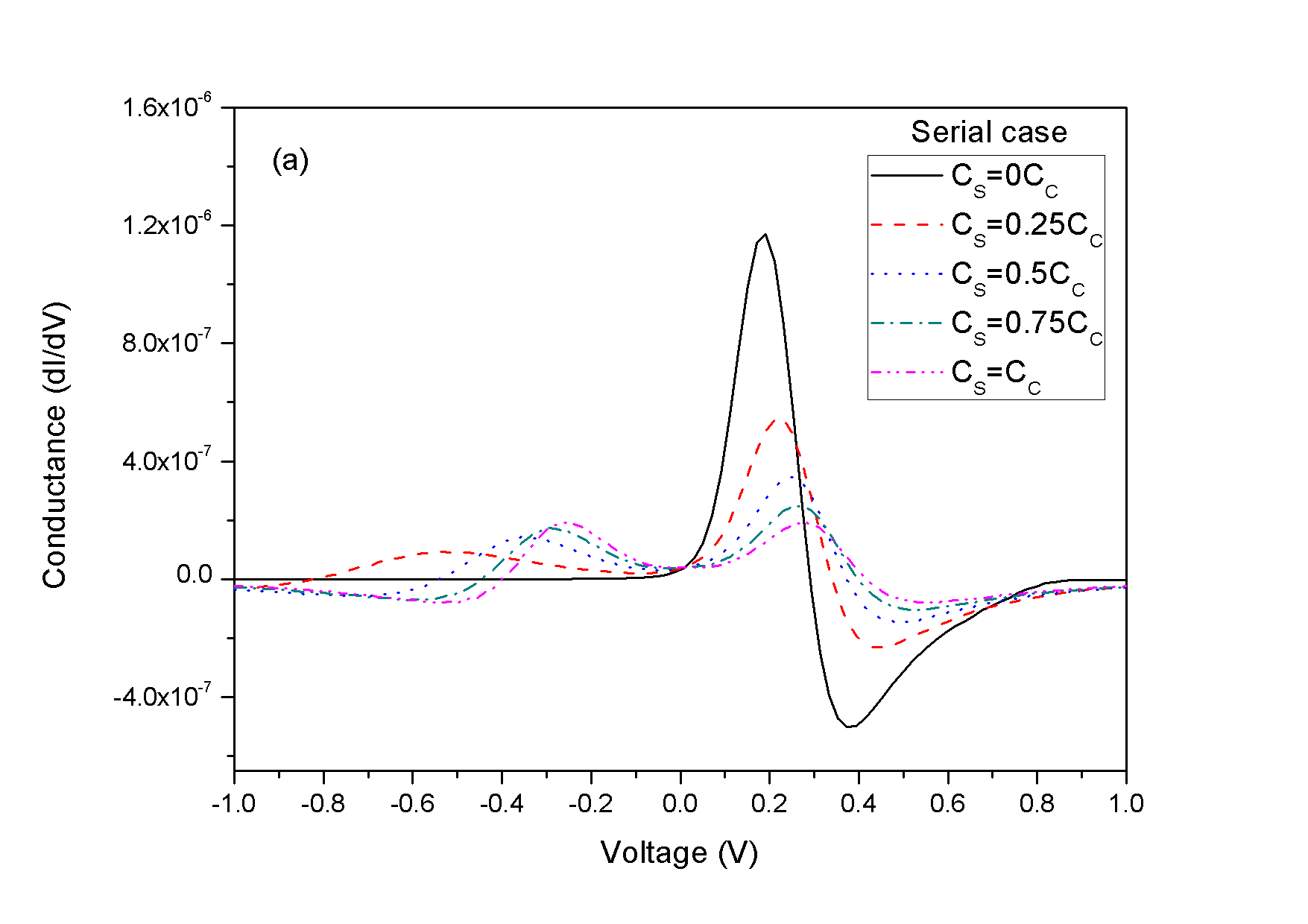}
\includegraphics[width=0.5\textwidth]{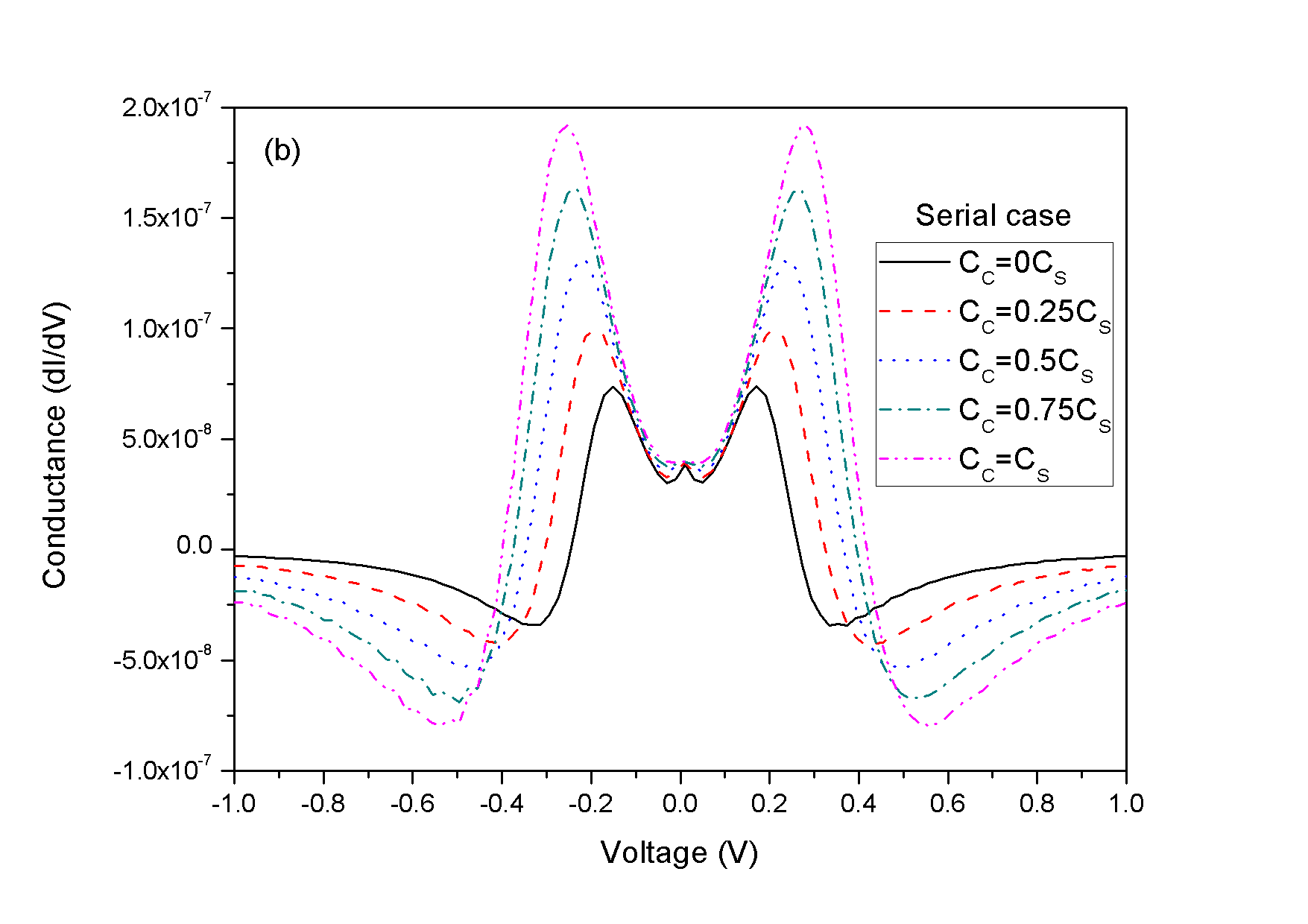}
 \caption{Conductance characteristics obtained in the serial case, in the same cases as before. NDR and rectification effect (a). Evolution of the resonant peak, maximum of conductance, and the NDR as a function of the coupling capacity between dots (b).}
\end{figure}

As we did before, the evolution of the energy levels inside each Qd taking into account (22) and (23) are 
\begin{equation}
\epsilon_1=\epsilon_0+\frac{C_s}{C_{tot}}(-qV_G)+\frac{C_c}{C_{tot}}U_2 \approx \epsilon_0+\frac{C_s+C_c}{C_s+2C_c}(-qV_G)
\end{equation}
\begin{equation}
\epsilon_2=\epsilon_0+\frac{ C_c}{C_{tot}}(U_1) \approx \epsilon_0+\frac{C_c}{C_s+2C_c}(-qV_G)
\end{equation}
Where we have used the same geometric recurrence as before. For simplification we have considered $C_s=C_{s2}$. Now the two conditions imposed before in order to transport occurs and can be summarized as $\mu_1>E1=E2>\mu_2$. As we can see from the expressions (30) and (31), $\epsilon_1$ increases faster than $\epsilon_2$ and thus the condition $\epsilon_1=\epsilon_2$ can not be accomplished. Introduced DOS broadening relaxes restrictive conditions for transport as a result of energy channel overlapping. When the voltage increases, the separation between $\epsilon_1$ and $\epsilon_2$ increases as well and then the overlapping between DOS decrease closing some channels and decreasing the outcoming flow. As a consequence a NDR appears due to the different electrostatic coupling in each dot \cite{20,21,22,23}. The $I(V)$ peak is related with the maximum overlapping between the DOS of the first and the second Qd \cite{24}. Moreover the width is related with the DOS broadening. 

\section{Conclusions}
A simple phenomenological model of the conduction between serial or parallel quantum dots has been developed based in rate equations for non coherent Qd system. This approach provides the most simple and transparent way for a description of electron transport. In the beginning we consider a system of two Qd linked by a ballistic channel and connected with the emitter and collector reservoirs. Level broadening has been introduced in the DOS changing the expression for current and the population inside each dot. Moreover this model can be extended easily to an array of N$\times$M Qd.\\
An important point has been the calculation of the local potential, solving the Poisson equation with appropriate boundary conditions for each dot and confi\-guration. Despite its simplicity the effect of self-charge has been token into account. As we may expect the calculation of the local potential inside each dot is the most important point. This local potential dominates de current characteristics since overlapping of the DOS is imposed in order to have electron transport. \\
I(V) and conductance curves have been obtained in a range of capacity values. NDR and I(V) resonant peak have been observed in the serial configuration due to the different electrostatic coupling.
In parallel configuration a conduction gap has appeared as some other authors have reported, therefore two threshold voltages have been obtained. Each Qd acts like an independent conduction channel and their electrostatic interaction is a second order effect. 

\section*{Acknowledgments}
This work was supported by NASCENT FP7-NMP-245977 European project. A.C. acknowledges support from ICREA academia program.


\begin{thebibliography}{XXX}
\bibitem{16}To solve the system we use the following procedure: we start with an initial guess for $U_{old}$, then calculate $N$ and the total potential $U_c$. Using the algorithm $U_{new}=U_{old} +\alpha(U_c-U_{old})$ we define a new potential $U_{new}$, where $\alpha<1$. The iterative process has to be repeated untill we find a $U_{new}$ which is sufficiently close to the original value. A simultaneous convergence of $U_1$ and $U_2$ is needed-.
\bibitem{19}The system can be described by Hamiltonian of the general form 
$$H=\sum_{k,\sigma} \epsilon_k^L a_{k,\sigma}^{\dagger}a_{k,\sigma}+\sum_{p,\sigma} \epsilon_p^R b_{p,\sigma}^{\dagger}a_{p,\sigma}+\sum_{\sigma} \epsilon_1 c_{1,\sigma}^{\dagger}c_{1,\sigma}$$
$$+\sum_{\sigma} \epsilon_2 c_{2,\sigma}^{\dagger}c_{2,\sigma}+U_1 n_{1 \uparrow} n_{1 \downarrow}+U_2 n_{2 \uparrow} n_{2 \downarrow}$$
$$+(\sum_{k,\sigma} V_k^L c_{1,\sigma}^{\dagger}a_{k,\sigma}+\sum_{p,\sigma} V_p^R c_{2,\sigma}^{\dagger}b_{p,\sigma}+\sum_{\sigma} V_d c_{1,\sigma}^{\dagger}c_{2,\sigma}+H.c)$$

Where, $ a_{k,\sigma}^{\dagger} ( a_{k,\sigma})$, $b_{k,\sigma}^{\dagger} ( b_{k,\sigma})$ y $c_{k,\sigma}^{\dagger} ( c_{k,\sigma})$ $i=1,2$  are the creation (annihilation) operators of an electron with the wave vector $k$ and spin $\sigma$ in the right lead, left lead and $i\!\!-\!\!th$ Qd respectively. $\epsilon_k^L$ and $\epsilon_p^R$ are the energies levels inside the leads,  $\epsilon_i\,\,(i=1,2)$ are the discrete energies levels of the $i\!\!-\!\!th$ dot. The fifth and sixth terms are the Coulomb integral for two electrons residing in the same dot. The last term describes electron tunnelling between the leads and dots.
\end{thebibliography}
\end{document}